We choose not to engage in any non-constructive polemics about the statement that we, the Russian participants of the project, did not discuss the contents of the article. We will simply enumerate some facts that disprove it.

Almost a year ago, even before the experiment itself was completed, our Western colleagues conducted a seminar discussing experiment results at the Physics Department of TUM E18. During this seminar, they demonstrated and discussed all the characteristic particularities of the spectra cited on Fig. 5 and 6 of our article. Immediately after the experiment was completed we wrote to our Western colleagues that we are processing the spectra, which constituted an invitation to join us in this effort. After we obtained the results last year, we prepared the article and placed it on the electronic preprint website, but received no comments regarding the content of the article from our coauthors. All of their objections regarded the list of coauthors.

J. Byrne formally is a participant of our project INTAS 00-00115a, coordinated by N. Severjins. We have also formally included J. Byrne into the coauthors list for our articles on radiative neutron decay for all the years that the project has been conducted. For example, in the article published in JETP Letters in 2002, he is listed as one of the first coauthors (could this possibly be the reason that this article's "scientific standard" does not provoke any doubts from the commentary authors, despite the fact that at this point we had only obtained a limit on the radiative mode, and in the current article we had already measured it!). But, throughout the five years that the project had lasted, he did not take any factual part in experiment proposals, preparation, conduction, or the processing of the obtained data. Due to these reasons, we excluded him from the list of –coauthors in our last experiment.

The proposal of this experiment and the equipment itself are Russian (that is not hard to prove using the published articles), so why do we have to give away the authorship to J. Byrne, especially given the fact that he had taken no part in the last experiment?!

We would like to emphasize that Fig. 5 and 6 of our articles present experimental spectra as they are, without preliminary data processing or background subtraction. The "forest of peaks" on Fig. 5 consists of just two main peaks (the peak of momentary coincidences and the neutron decay peak itself), which were repeatedly observed both by our group while we were measuring the radiative decay mode and by another (of which one of us, V.A. Solovei, was a participant) while measuring the asymmetry of beta-electron emission.

The remaining peaks are small, with just seven peaks distinct from the statistical fluctuations. These occurred because of the noises in electric circuit of the FRMII neutron guide hall. There are no other physics-related reasons for their occurrence. These peaks appeared and disappeared depending on the time of day, reaching their maxima during the work day and disappearing over the weekends. Such behavior was observed throughout the experiment as we collected statistics. Since the nature of these seven small peaks is in no way related to radiative decay, we did not emphasize them in our article.

Three peaks are distinct on Fig. 6. The problem is that the leftmost radiative peak in channel 103 is observed on non-homogenous double-humped background. This non-homogenous double-humped background can be obtained from Fig. 5 spectrum with the response function (see functional equation (1) of the article). Using this method, one can confidently define the background and so obtain radiative events located in the leftmost peak in channel 103. At the same time, the narrow peak with the maximum in channel 106 is the response to the narrow peak of false coincidences in channels 99-100 on Fig. 5, and the second peak in this double-humped background on Fig. 6 is the response to the peak in channels 117-127 on Fig. 5.

Of course, the width of these response peaks is greater than the width of the two original peaks on Fig. 5, the width of the narrow peak increases more, the distance between the peaks themselves diminishes, and the narrow peak moves to the right towards the wider peak more than this wide peak moves in opposite direction to the left . Such behavior is described using the standard method of response function, and there is nothing unusual about it. Here it's also appropriate to remark on another peculiarity observed throughout the experiment and which emphasizes the physics-related nature of the peak in channel 103 on Fig. 6. As noted above, the noise peaks on Fig. 5 were not stable, and neither were the small peaks in channels 96 and 115 on Fig. 6. Besides, if the small noise peaks in the spectrum of double coincidences disappeared, the small peaks in channels 96 and 116 in the spectrum of triple coincidences disappeared as well. The radiative peak in channel 103 was stable throughout, it never "migrated" to a different channel and its growth was stable, regularly collecting the same number of events during the same stretch of time!

As for the wide, almost indistinguishable peak in channel 165 on Fig. 6, its influence on radiative peak in channel 103 is negligible. Its nature is in no way related to the researched phenomenon, so we do not discuss it in our article.

It follows that we remain firm in the position related in the article. We measured the radiative decay mode and gave the value of its relative intensity ( B.R.=$(3.2\pm1.6)\cdot10^{-3}$ with CL=99.7% and lower energy limit of radiative gamma-quanta measured equal to 35 KeV). The radiative events of the triple coincidences spectrum were distinguished using the response function method. All the particularities of the non-homogenous background, against which we see this radiative peak in channel 103 on Fig. 6, are explained using this method. At the same time, it's important to note that the average B.R. value exceeds the theoretical value, calculated within the framework of the standard electroweak theory. It would only be possible to check this deviation with the help of the next experiment, in which we could realistically obtain an error of less than 10% versus the current 50%.

We understand the full extent of our financial dependence on the Western collaborators, and, what's worse, currently in Russia there are no active sources of neutron beams that would have suitable intensity for conducting our experiment. Despite this, we cannot repudiate the result we cited in our article. Of course, it should be clear that a significant part of the work conducted (i.e. computer Monte-Carlo simulation of the experiment) was not reflected in our short article. However, all the major central points and the final results of the work are presented there.

In conclusion, we cannot ignore the provocatively rude attitude of the commentary, which does not contain any scientifically based constructive criticism, much less any alternative suggestions.

Leader of the Russian Research  
Center "Kurchatov Institute" team  
in INTAS project 00-00115a          R. U. Khafizov

Leader of PNPI team in INTAS  
project 00-00115a          V.A. Solovei